\documentclass{aa}
\usepackage{psfig}
\usepackage{jour}
\usepackage[round]{natbib}
\bibpunct{(}{)}{;}{a}{}{,}
\newcommand{\VEV}[1]{\left\langle #1\right\rangle}
\begin{document}

\title{Error estimation for the MAP experiment}

\author{
S.~Prunet\inst{1} \and R.~Teyssier\inst{2} \and
S.~T.~Scully\inst{3} \and F.~R.~Bouchet\inst{4} \and
R.~Gispert\inst{5}
}
\institute{Canadian Institute for Theoretical Astrophysics,
McLennan Labs, 60 St George Street, Toronto ON M5S 3H8, Canada
\and CEA/DSM/DAPNIA Service d'Astrophysique, Centre
d'Etudes de Saclay, Bat. 709, 91191 Gif/Yvette Cedex, France
\and Department of Physics and Astronomy, Valparaiso University, Valparaiso, IN 46383, USA \and
Institut d'Astrophysique de Paris, CNRS, 94 Bld
Arago, 75014 Paris, France
\and Institut d'Astrophysique Spatiale, CNRS-Universit{\'e}
Paris-Sud, Bat. 121, 91405 Orsay, France}

\offprints{S.~Prunet,
\email{prunet@cita.utoronto.ca}}
\abstract{
We report here the first full sky component separation and CMB power
spectrum estimation using a Wiener filtering technique on simulated data from 
the upcoming 
MAP experiment, set to launch in early
2001. The simulations included contributions from the three dominant astrophysical
components expected in the five MAP spectral bands, namely CMB
radiation, Galactic dust, and synchrotron emission. We assumed a
simple homogeneous and isotropic white noise model and performed our
analysis up to a spherical harmonic multipole $\ell_{\rm max}=512$ on the
fraction of the sky defined by $|b| > 20^o$. We find that the
reconstruction errors are reasonably well fitted by a Gaussian with an
rms of 24 $\mu $K, but with significant deviations in the tails. Our
results further support the predictions on the resulting CMB power
spectrum of a previous estimate by \cite{bouchet99a}, which entailed a
number of assumptions this work removes.
\keywords{methods: data analysis -- cosmic microwave background}
}
\markboth{Error estimation for the MAP experiment}{Prunet {\emph et al./}}
\date{}
\maketitle

\section{Introduction}

The upcoming CMB satellite experiments MAP and the Planck Surveyor offer
an unprecedented opportunity to measure CMB temperature fluctuations
on the whole sky.  A major hurdle in extracting the primary CMB signal
from data, apart from noise, is the removal of the Galactic and extragalactic
foregrounds. However, as the foregrounds differ from CMB in both
frequency dependence and spatial distribution, one can reduce their
residual level in a multi-frequency CMB experiment. A multi-frequency,
multi-resolution, Wiener filtering method to optimally extract the CMB
component was developed \citep{bouchet96,teg,bouchet99a} and lead to
detailed predictions for the accuracy achievable with several
experiments, and in particular, for MAP \citep{bouchet99a}. These
papers were purely semi-analytical and entailed a number of simplifying
approximations and assumptions. A practical implementation was performed 
by \citet{bouchet96} on the simulated data produced by \citet{bouchet95}.
These numerical studies
confirmed that the residual contamination after cleaning the map is
much smaller than the CMB primary signal and therefore the
foregrounds may not be a major obstacle in the extraction of CMB
temperature angular power spectrum if the real sky behaves similarly
to the assumed model.

These simulation studies were based on filtering the data obtained on small
patches of the sky of typically $10^o$ by $10^o$ angular size. They did
not involve the assumption of Gaussian foregrounds since actual
templates were used ({\it e.g.} portions of the Haslam and IRAS maps at
respectively 408 GHz and 100 $\mu $m). The results, however, could not be
safely extrapolated to the full sky as a result of three facts:
1) the Wiener filter used
was always optimal for the particular (small) region being analyzed
instead of being an average filter for all the sky; 2) nothing could be
said about modes larger than the map size; 3) there was no really
satisfactory way of co-adding individual map results even if all maps
covering the sky had been used (one problem being the use of Fourier
transforms with periodic boundary conditions). Note that a similar
approach based on a multi-frequency maximum entropy method
in Fourier space on the same simulations leads to comparable
optimistic conclusions (\citet{hobson98}, applied to simulated MAP
data by \citet{jones99}), but with similar caveats.

In this work, we extend the Wiener filtering method to full-sky maps
with a symmetric Galaxy cut ($|b| > 20^o$) and discuss our solutions
to the various technical difficulties encountered. We report the results
for the MAP experiment and compare them with the relevant predictions
of \cite{bouchet99a}. A companion paper will detail the implementation
and give results for the Planck Surveyor experiment that requires
using a much more sophisticated foreground model due to the higher
resolution and higher sensitivity of that experiment.

\section{Full Sky Wiener Filtering}

Given a set of full sky maps observed in different frequency bands
(at 22, 30, 40, 60 and 90 GHz for the MAP experiment), we want to recover
the underlying CMB signal by combining the channels in an optimal way,
that takes into account both the foreground contaminants and the instrumental
noise.

We write the signal as the sum of the true data and the noise in each
band, $\mathbf{d}_{\nu} = \mathbf{s}_{\nu} + \mathbf{n}_{\nu}$, where
each vector component is a pixel map. The filtered signal,
$\mathbf{\hat{s}_{\nu}}$, is a linear combination of the data,
$\mathbf{\hat{s}_{\nu}}=W_{\nu\mu } \mathbf{d_\mu }$, where $W$ is the
linear filter used in the multi-frequency reconstruction. The Wiener
filter is specifically obtained by minimizing the trace of the
covariance matrix of the error map defined as $\mathbf{e_{\nu}} =
\mathbf{\hat{s}_{\nu}} -\mathbf{s_{\nu}}$. We get the usual formula $W
= S (S + N)^{-1}$ where $S = \VEV{\mathbf{s} \mathbf{s}^T}$ is the
covariance matrix of the true signal and $N$ is the covariance matrix
of the noise. The noise characteristics should be defined as precisely
as possible for a given experiment, but the signal is {\it a priori}
unknown. The {\it prior probability} given by the covariance matrix
$S$ is a key ingredient in the Wiener filtering method. We assume here
that the signal in a given frequency band is a linear combination of
several astrophysical processes including the CMB, the different
Galactic emission processes, and other extragalactic foregrounds.
This can be formulated as $\mathbf{s_{\nu}}=A_{\nu p}\mathbf{x}_p$
where $\mathbf{x}_p$ is a reference template for a given astrophysical
process. This assumption allows us to factorize the spectral and
spatial properties of the signal. The observation matrix $A$ can also
account for the beam as a convolution operation on the input maps. The
covariance matrix of the signal in the different frequency bands is
readily obtained by $S = A C A^{T}$ where $C$ is the covariance matrix
of the templates (for uncorrelated processes, it is block diagonal).
The last step in the component separation is to recover from the
filtered data the estimates of the astrophysical processes by a
Least-Square fit to the recovered signal, namely
$$
\mathbf{\hat{x}}_p =
(A^{T} A)^{-1} A^{T} \mathbf{\hat{s}_{\nu}} 
= C A^{T}\left(A C A^{T} + N\right)^{-1} \mathbf{d_{\nu}},
$$
which is the formula for multi-frequency Wiener filtering \citep{bouchet96}.

One problem in analyzing full sky maps is the huge number of pixels
that one must deal with. A direct pixel-based approach of the
Wiener filtering technique is far beyond the capabilities of current and near-future
supercomputers. One possibility is to work with coefficients of a
harmonic decomposition where convolutions (of the template by the
optical beam) translate into mere multiplications. Indeed previous
simulation analyses all dealt with Fourier coefficients. For the full
sky analysis presented here, we use instead spherical
harmonics coefficients. This harmonic basis change is particularly
useful if each astrophysical process can be well approximated by a
homogeneous and isotropic random field with a given power
spectrum. In this case, the Wiener filter can be computed for each
multipole $\ell$ independently.

Here we consider that, to first order, the Galactic emission can be
modelled as the sum of spatial templates for each astrophysical emission
mechanism ({\it i.e.} dust, synchrotron) multiplied by their associated frequency
dependence.  At high galactic latitude this appears to be a good
approximation \citep[see {\it e.g.}][and references therein]{bouchet99a}. 
At low galactic latitude, the situation is more complex as
strong variations of the effective spectral index are observed in
the Galactic disk. In addition, the presence of very bright sources at
low latitude strongly violates the assumption of Gaussianity that is
required for Wiener filtering to be linearly optimal, while this assumption
is more reasonable at high latitude. Moreover, the presence of
these bright, highly localized features is likely to ruin any component separation
analysis that assumes statistical isotropy.
All these points, together with the fact that CMB emission is dominant
only outside the Galactic plane, indicate that a careful removal of
the Galactic plane region is necessary in the analysis of full sky CMB data.

\section{Working with a Galaxy Cut}

\subsection{Building a New Orthonormal Basis on the Cut Sky}

The problem that arises as soon as one removes the Galactic plane
region from a full sky map is the loss of orthonormality of the
spherical harmonics basis \citep{gorski94}. As was pointed out in
the last section, a Wiener filter results from minimizing the trace of the
covariance matrix $E$ of the error map. If we write for a given map
$\mathbf{x}$ that $\mathbf{x}=Y\mathbf{\tilde{x}}$, where $Y$ are the
spherical harmonics ($Y_{lm}$'s) and $\mathbf{\tilde{x}}$ are the map's
multipoles, we get
$$
Tr(E) = \langle\mathbf{e^{\dagger}}\mathbf{e}\rangle =
\langle\mathbf{\tilde{e}^{\dagger}}Y^T Y\mathbf{\tilde{e}}\rangle
\neq  \langle\mathbf{\tilde{e}^{\dagger}}\mathbf{\tilde{e}}\rangle
$$
where $\mathbf{e}$ is the error map of any given process.  Therefore,
Wiener filtering the data in harmonic space is not equivalent anymore
to Wiener filtering the data in real space.  The usual way to solve
this problem is to build a new orthonormal basis on the cut
sky. \citet{gorski94} proposed a method for the COBE-DMR data based
on a Cholesky decomposition of the coupling matrix $Y^T Y$ of the
$Y_{lm}$'s.
In the course of our work, we found that this method works well for
$\ell_{\rm max}$ up to 40 - 50, but the Cholesky decomposition failed to
converge for higher multipoles because the coupling matrix
becomes ill-conditioned due to numerical truncation errors. We
have therefore used instead an orthonormalisation scheme based on the
Singular Value Decomposition of the $Y$ matrix.  The SVD can be written $Y =
UDV^{T}$ where $U^{T}U=I$, $V^{T}=V^{-1}$, and $D$ is a diagonal matrix
containing the singular values. This decomposition is numerically stable. 
The new basis is
given by the matrix $U$ and the decomposition becomes
$\mathbf{\tilde{x}} = U^T \mathbf{x}$ (if $\mathbf{x}$ can be
represented on that basis, {\it i.e.} $\mathbf{x} = U \mathbf{\tilde{x}}$,
then $\VEV{\mathbf{x}^T\mathbf{x}} =
\VEV{\mathbf{\tilde{x}}^T\mathbf{\tilde{x}}}$ as desired). Vectors of
the new basis that correspond to vanishing singular values can be
dropped as their support is confined to the Galactic cut.
Note that for high resolution maps, this new basis can be
obtained at a reasonable computational cost (but {\it only} 
if the cut is symmetric).
We therefore restrict ourselves to a symmetric, isolatitude Galaxy cut
(with $|b| > 20^o$ for example).

\subsection{Power Spectrum Estimator on the Cut Sky}

One important ingredient of the Wiener filtering technique is the covariance
matrix of the astrophysical processes. One possibility is to give an
analytical form of the $C(\ell)$ as a prior. In practice, however, it is more
useful to compute the power spectrum directly on a first estimate
of each astrophysical template \citep[obtained for instance with an SVD 
method, see][]{bouchet99a} and ultimately, do an iterative
Wiener filtering by refining the estimate of the power spectra at each
step. In this paper, we compute directly the power spectrum of the
templates used in the simulated data, but by using only the pixels at
high galactic latitudes. \citet{tegmark95} described a method to obtain
an optimal power spectrum estimator for each $\ell$ on an incomplete
sphere by minimising the leakage between
multipoles \citep[see][for details]{tegmark95}. We use this method to
design ``optimal masks'' or ``apodising functions'' to compute the
$C(\ell)$ on our Galaxy cut templates.  Our implementation was designed to work
in the framework of the orthonormal basis on the cut sky described
in the last section.  An intermediate possibility would be to exploit the
smoothness of the different power spectra. 
The computation of the power spectra in terms of
band-powers \citep{bond00} might offer in this respect a good
compromise between spectral resolution and computational speed, but we
did not test it.

\section{Numerical Techniques}

In this work, we use extensively the HEALPix pixelisation
scheme\footnote{see {\rm http://www.eso.org/~kgorski/healpix/}}.  We
worked at a resolution ${\rm n_{side}}=256$ with a pixel size of 13.7
arcmin\footnote{Healpix divides the sphere in $12$ main regions, each
being hierarchically divided in ${\rm n_{side}^2}$ pixels}. This
pixelisation choice is a little too coarse to extract all the CMB
information from MAP since its 90 GHz channel has a 12.8 arcmin FWHM
beam. As we shall see, however, it is enough to capture the essential
features of the Wiener filtering possibilities for the MAP experiment.

\begin{figure}
\centerline{\psfig{file=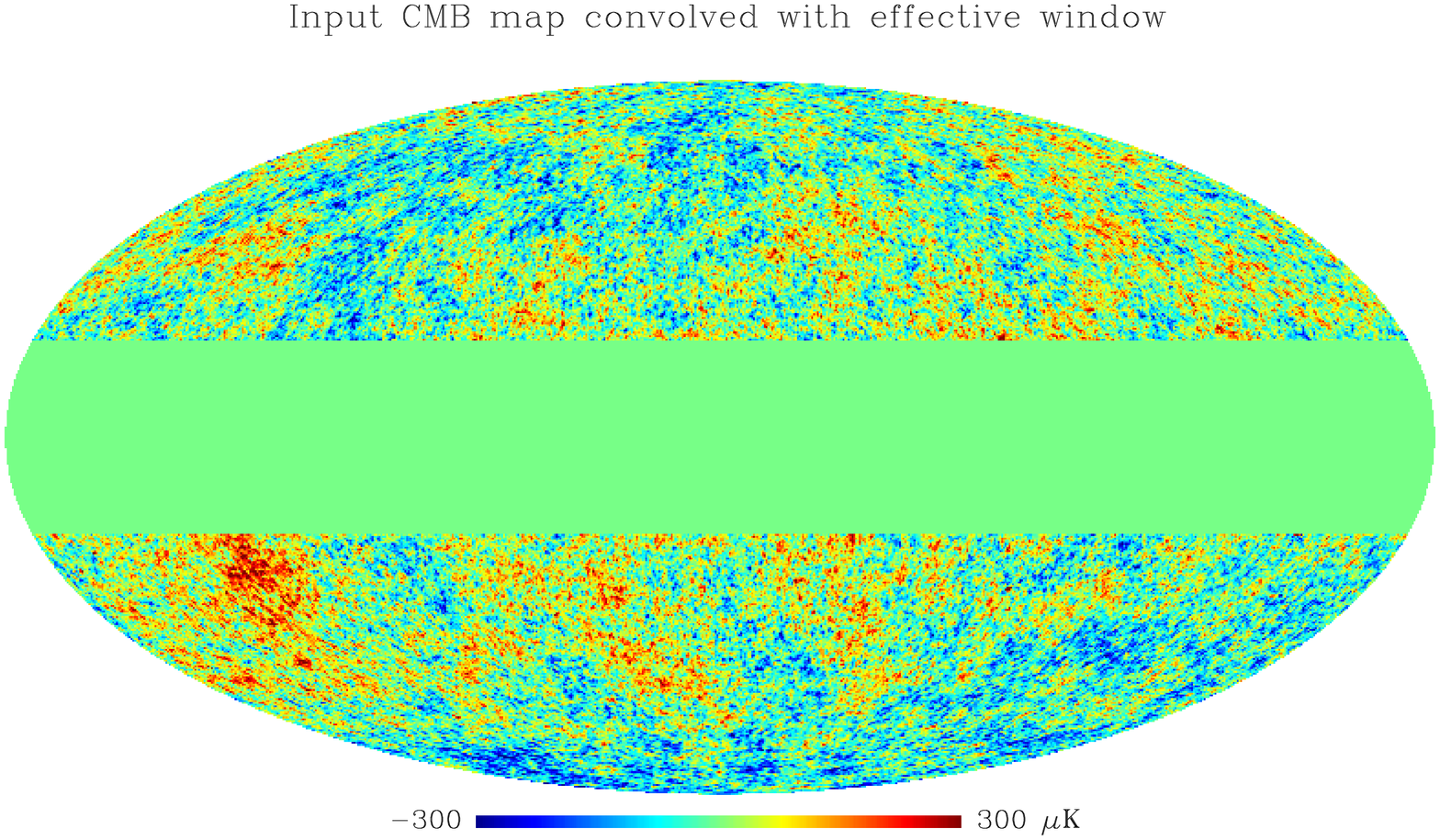,width=9cm,angle=90}}
\centerline{\psfig{file=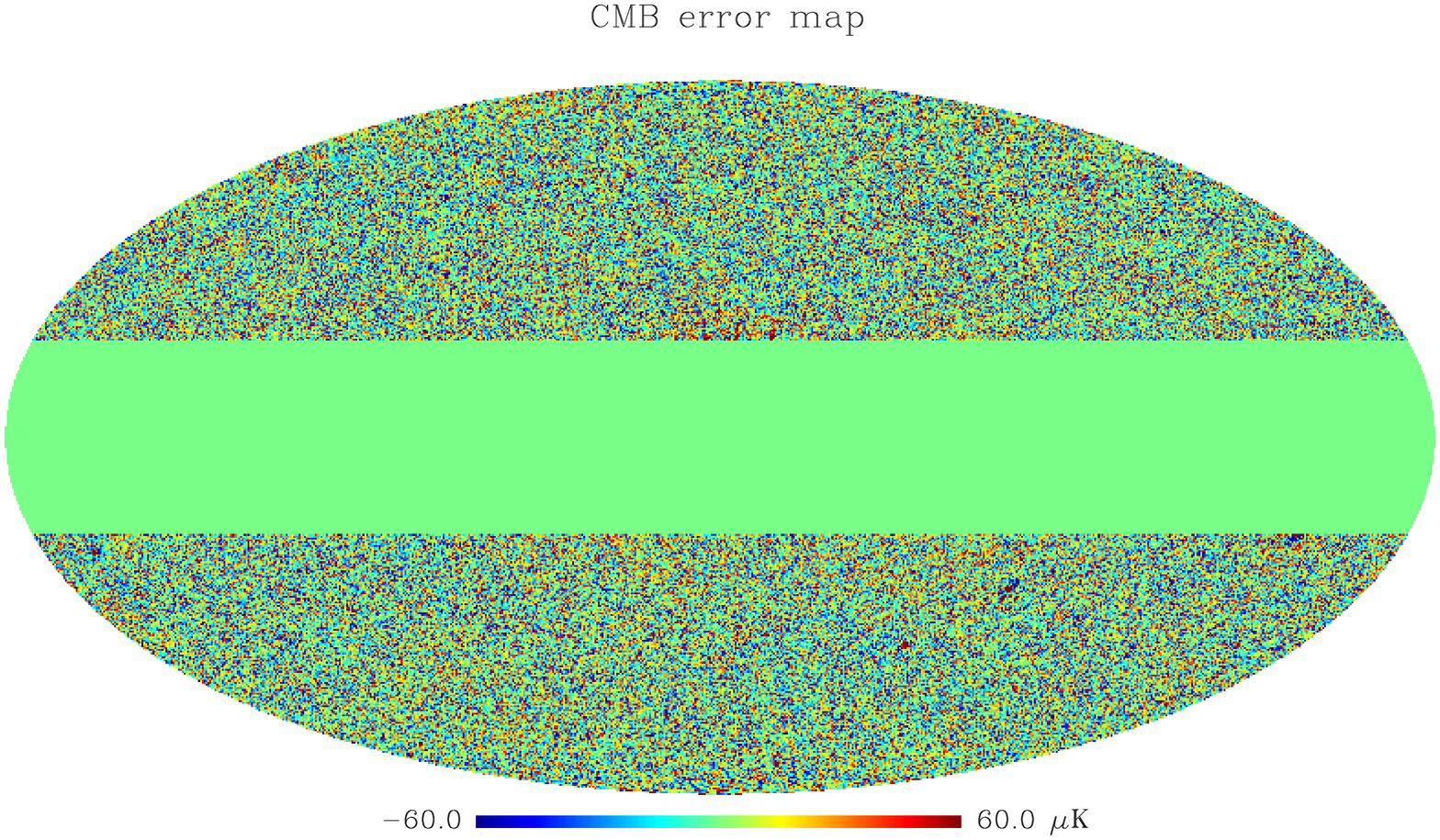,width=9cm,angle=90}}
\caption{{\em Upper panel}: The CMB input map, convolved with the
effective window function of the experiment. {\em Lower panel}: The
CMB error map. Note the different scale used here.  The error is
mostly white noise, except near the galactic cut.  Several bright
spots are also visible.}
\label{fig2}
\end{figure}

Due to the constraint that our Galaxy cut is symmetric with respect
to the Galactic plane, we can address each $m$ for odd and even
$\ell$ separately. This rotational symmetry reduces the computational
complexity from $O({\rm N_{pix}^2})$ down to $O({\rm N_{pix}^{3/2}})$.
Computing the
new basis represents $20$ minutes on a single processor of a SGI O2000
parallel system for ${\rm n_{side}}=256$ and $\ell_{\rm max} = 512$.
Recomputing the new basis each time we need to analyze a map is not
very efficient. We therefore store permanently the new coordinates,
which requires approximately 1 GB space in double precision.
Computing the ``optimal masks'' for power spectrum estimates is much
more computationally expensive. The core of the method is
to solve a General Eigenvalue Problem ($A x =\lambda B x$) 
\citep{tegmark95}, and to
compute only the first eigenvector. This process takes $45$ minutes 
on the same parallel computer using $10$ processors
running in parallel under MPI. The storage of these ``optimal masks''
also requires approximately 1 GB in double precision. Note that storage
requirements are important here, since extrapolating our results to
${\rm n_{side}} = 1024$ and $\ell_{\rm max} = 2048$ leads to a total
computational time of $48$ hours with $10$ processors and a total
memory requirement of $128$ GB. We have, however, identified possible ways to
decrease the CPU and memory requirements
each by a factor of $4$.

\section{Results and Discussion}

\begin{figure}
\centerline{\psfig{file=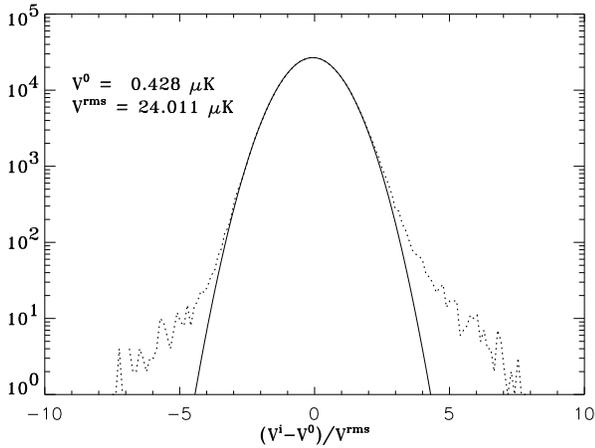,width=9cm}}
\caption{Distribution of the CMB map reconstruction error (dotted line),
obtained by subtracting the recovered CMB map from the input CMB map
smoothed with the effective window function for the CMB (``quality
factor'', see \citealt{bouchet99a}) produced by Wiener filtering (see text).
Also shown is the best-fit
Gaussian (solid line) with its mean and rms. Note that the non-zero mean value 
shows the small level of contamination of the reconstructed CMB by the foregrounds
monopoles. It will however not be an issue for the real observations since MAP is
a differential experiment. }
\label{fig1}
\end{figure}
\begin{figure}
\centerline{\psfig{file=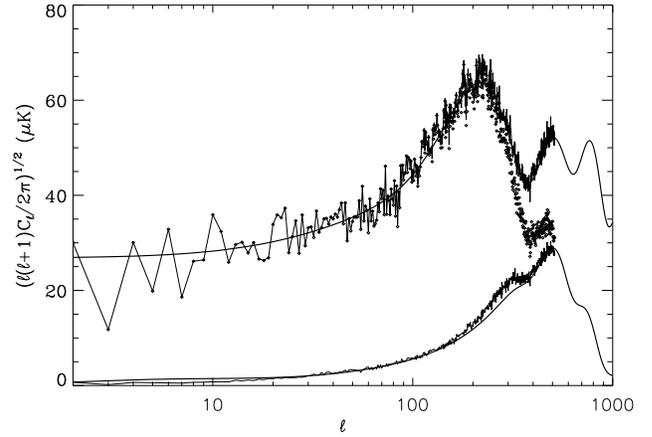,width=9cm}}
\caption{Comparison of the input sCDM angular power spectrum (upper
thick solid line) to that of the reconstructed (Wiener-filtered, hence
smoothed) CMB map (dots, middle noisy curve). The latter can be used to
obtain an {\emph unbiased} estimate of the input spectrum (thin solid
line, on top of the input spectrum) by using the quality factor
produced by the analysis (see text). The lower (thick solid) curve shows the
semi-analytical prediction of the spectrum of the
reconstruction error by \citet{bouchet99a}, together with the actual
estimate (thin solid) from our full-sky error map.}
\label{fig3}
\end{figure}

In this work we modelled the Galactic emission to be composed only of
synchrotron and dust emission. Other forms of emission such as uncorrelated 
free-free emission or point and SZ sources were left out for simplification of the 
interpretation of the results. We will address those additional components 
in a future paper. The dust emission was simulated
using the Finkbeiner et al. DIRBE/IRAS composite $100 \, {\rm \mu 
m}$ map \citep{finkbeiner}, and the synchrotron emission using the
Haslam $408\,{\rm MHz}$ map \citep{haslam} with spectral index $-0.9$. The dust
was assumed to have a single $18\,{\rm K}$ component with emissivity index
of $2$, together with a correlated free-free like component of spectral index 
$-0.15$. 

Figure~\ref{fig2} shows the input CMB map, smoothed at the expected
resolution of the reconstructed map, and the error map, which is the
difference of the input CMB map and the recovered one. As the latter
shows, the overall reconstruction is excellent, although a few
glitches associated with low galactic latitude HII clouds are visible
close to the Galaxy cut. This justifies {\it a posteriori} the removal of
the Galactic plane from our analysis. Figure~\ref{fig1} gives the
distribution of the reconstructed errors, whose bulk is well
approximated by a Gaussian distribution with an rms of 24 $\mu $K. Note
though the low level wings that are likely associated with the
residual imprint of HII clouds. The corresponding pixels would
probably have to be dropped out from any ensuing analysis.
 
An important feature of Wiener filtering is that the quality 
factor(s)\footnote{The
  quality factors are the generalized instrumental window functions
  {\em for each recovered astrophysical process}, taking into account
  the angular resolution and the detector noise of every channel, as
  well as the contamination of other processes, see
  \citealt{bouchet99a}.} of the instrument
is an output of the analysis. The expected
resolution of the reconstructed map is thus known directly and can be
accounted for in later stages of the statistical analysis. Indeed
we used it here to convolve our input CMB map to the expected
resolution of the recovered map.

For cosmological purposes, the most important statistic is the
angular power spectrum of the CMB. We therefore show the power
spectrum of the recovered CMB maps together with the spectrum of the
reconstruction error in figure~\ref{fig3}. The spectra are computed
{\it for each $\ell$} using the minimal leakage apodising masks on the
incomplete sphere (see section 3.2). The curves abruptly end at
$\ell=512$, the resolution limit of the present analysis, where
incidentally the error is of the order of the CMB signal.  Note
however that a box car averaging of the $C(\ell)$ with $\Delta \ell$
around 10 will reduce the noise level, and therefore a strong (and
thus useful) cosmological signal is still expected above $\ell=512$,
likely up to $\ell=1024$. We will extend this work in a companion
paper to a spatial resolution of at least $\ell=1024$.  The main
conclusion of the current work is that we confirm the semi-analytical
predictions of \citet{bouchet99a}, but using this time {\it a full sky
analysis} of simulated MAP data.

\acknowledgements
Simulations and map analysis were performed on MAGIQUE, an SGI
02000 14 processors parallel system at IAP. The authors would like
to thank Didier Vibert for his help on this project.


\bibliographystyle{apj}
\bibliography{ms7}

\end{document}